# Color Image Encryption Based on Chaotic Block Permutation and XOR Operation

Raneem A. Aboughalia
Department of Electrical and Electronic Engineering
University of Tripoli
Tripoli, Libya
Raneem.abg@gmail.com

Osama A. S. Alkishriwo
Department of Electrical and Electronic Engineering
University of Tripoli
Tripoli, Libya
Alkishriewo@yahoo.com

*Abstract*— In this paper, chaotic block image permutation and XOR operation are performed to achieve image encryption. The studied method of encryption makes use of chaotic systems properties for secure and speed image encryption. Firstly, the original image is divided into blocks of equal size. Then, two chaotic maps are used to generate two key streams which are permuted among themselves to produce one key steam. The image blocks are then shuffled using part of key stream. Finally, scrambled image is diffused by XOR operation with the key stream to get the encrypted image. The experimental results of several performance analyses about the pixel correlation, various statistical analysis, information entropy analysis, differential analysis, the key space and key sensitivity analysis, show that the algorithm can resist several know attacks effectively and has the advantages of large key space, high security, and high speed, assuring safety performance and secure image encryption.

## I. INTRODUCTION

Image information transmission has increased rapidly with the incredible development of internet technologies and wireless communication networks. All kinds of multimedia data like digital image, audio, text and video can be accessed easily on internet. As a result, cryptographic techniques are required to accomplish a certain level of confidentiality and information security, during data storage and transmission.

There are a number of data protection approaches including a wide variety of algorithms used for data encryption. However, traditional encryption algorithms which are proposed for text information such as DES, IDES, RSA are not suitable for image encryption because of the larger scale of data, higher redundancy and stronger correlation of pixels in images [1].

In literature, various techniques have been developed to design secure image encryption algorithms. Among these approaches is the chaos image encryption schemes which have many excellent properties, such as ergodicity, random-like behavior, and sensitivity to initial values, etc, which can be used as an ideal image encryption method. Chaotic characteristics attracted the attention of cryptographers. They applied chaos theory to cryptography and have developed many image encryption based on chaos. Compared with traditional encryption algorithm, image encryption chaos-based has large key space, which is implemented simply and is robust.

As early in [2], Matthew proposed a logistic map based encryption algorithm. In [3], Fridrich firstly introduced permutation-diffusion architecture for chaotic image encryption. Later, many researchers paid a great attention to chaotic image encryption technology and proposed a variety of algorithms [4-7].

Here in this paper the emphasis is on chaotic based color image encryption algorithm, employing two different chaoticmaps to generate two key streams X and Y that areused to scramble image blocks and to perform the bitwise XOR operation to achieve image encryption.

The remainder of this paper is organized as follows. In Section II, chaotic maps are briefly addressed. In Section III, the algorithm will be described in details. In Section IV,simulation results and security analysis are presented.In Section V, comparison between the proposed algorithm test results with other algorithms is given. The conclusions are given in Section VI.

## II. CHAOTIC MAPS

### A. Logistic map

Logistic Map is a polynomial equivalence of degree 2.The map was popularized in a seminal 1976 paper by the biologist Robert May, in part as a discrete-time demographic model analogous to the logistic equation first created by Pierre François Verhulst. Mathematically, the logistic map is written as:

$$x_{i+1} = \mu\, x_i\, (1 - x_i) \qquad (1)$$

where, $\mu$ and $x_0$ are the control parameter and initial state value, respectively. If one choose $\mu \in [3.57, 4]$, the system is chaotic.

## B. Duffing map

The Duffing map (also called as 'Holmes map') is a discrete-time dynamical system. It is an example of a dynamical system that exhibits chaotic behavior. The Duffing map takes a point $(x_n, y_n)$ in the plane and maps it to a new point given by (2):

$$x_{i+1} = y_i$$
$$y_{i+1} = -b\,x_i + a\,y_i - y_i^3 \qquad (2)$$

The map depends on the two constants *a* and *b*. These are usually set to *a*= 2.75 and *b*= 0.2 to produce chaotic behavior. It is a discrete version of the Duffing equation.

## III. ENCRYPTION AND DECRYPTION ALGORITHM

The studied image encryption algorithm uses block permutation and XOR operation. Initially the original color image is divided into blocks of $m \times m$. Chaotic maps are being used to generate key stream, with initial condition $(x, y, v, w)$. The generated key stream is used to permute the divided image blocks. The shuffled image blocks of images are merged to form a single image. Fig.1 shows the block diagram of the encryption method.

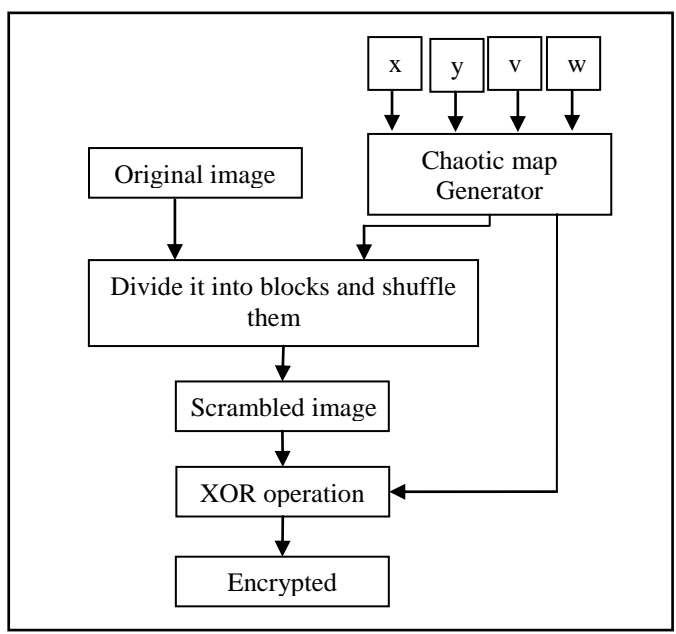

Fig. 1. Block diagram of encryption algorithm.

### A. Encryption Algorithm

The following are the sequence of steps used to encrypt images. Inputs to the algorithm are the original images, desired blocks size, initial condition generated using random function, chaotic map key stream, while the Output is the encrypted image.

Step1: Input the original image of size $(n \times n)$, and the block size.
Step2: Generate random initial conditions $(x, y, v, w)$, taking into account the original image pixel values.
Step3: Using the initial conditions generate chaotic maps key streams $X$ and $Y$.
Step4: Original image is divided into blocks of size $(m \times m)$.
Step5: Shuffle the block of images using part of the key stream of length $(n/m)^2$.
Step6: Combine the scrambled image blocks into a single image.
Step7: Perform XOR operation between pixel data of the scrambled image and the key stream generated by chaotic maps.

## IV. SIMULATION RESULTS AND PREFORMANCE ANALYSIS

In this section, experimental results and performance analysis of the studied image encrypting algorithm is presented. Test images were chosen to be $512 \times 512 \times 3$ standard 8-bit color images.

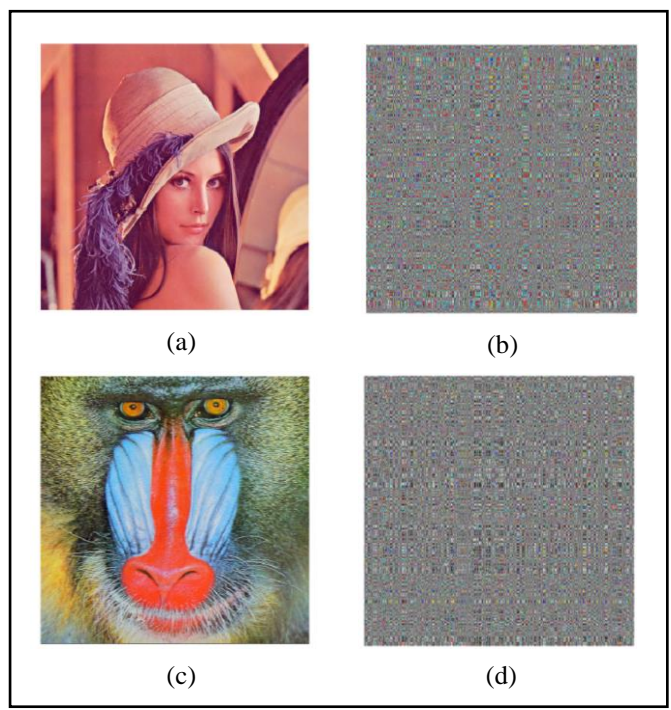

Fig. 2. Encryption results: (a) plain text Lena image, (b) encrypted Lena image, (c) plain text Mandrill image, (d) encrypted Mandril image.

Figure 2 shows the encryption results of different test images using different chaotic maps, while Fig. 3 shows the encryption and decryption process steps of the proposed encryption algorithm.

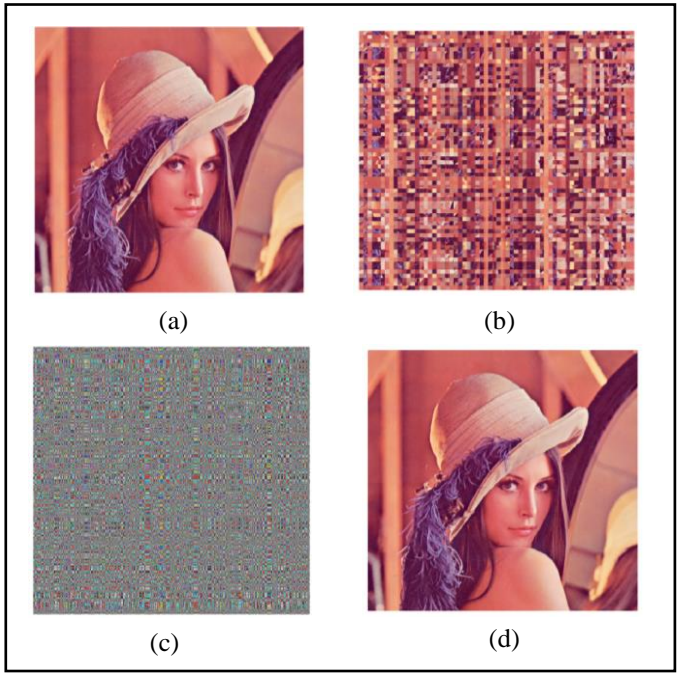

Fig. 3. Image encryption and decryption process: (a) plaintext image, (b) permutated image, (c) encrypted image, (d) reconstructed image.

A. *Differential Attack*

In general the two most common measures used to quantify this requirement are the Number of Pixel Change Rate (NPCR) and the Unified Averaged Changed Intensity (UACI), which are used to evaluate the strength of image encryption algorithms with respect to differential attacks.

If $C_1$ and $C_2$ are two different encrypted images whose corresponding plaintext images differ by only one pixel, then the NPCR and UACI can be mathematically defined as

$$\text{NPCR} = \sum_{i,j} \frac{D(i,j)}{M \times N} \times 100\% \quad (3)$$

$$\text{UACI} = \frac{1}{M \times N} \sum_{i,j} \frac{|C_1(i,j) - C_2(i,j)|}{255} \times 100\% \quad (4)$$

where,
$$D(i,j) = \begin{cases} 0 & if \ C_1(i,j) = C_2(i,j) \\ 1 & if \ C_1(i,j) \neq C_2(i,j) \end{cases} \quad (5)$$

From the NPCR and UACI results illustrated in Tables 1 and 2, we can find that our proposed algorithm is very sensitive to small changes in plain text images

| Image | Red | Green | Blue |
|---|---|---|---|
| Peppers | 99.5953 | 99.6044 | 99.6201 |
| Lena | 99.6071 | 99.6246 | 99.6208 |
| Sailboat | 99.6258 | 99.6197 | 99.6174 |
| Mandrill | 99.6193 | 99.6082 | 99.6227 |
| Airplane | 99.6048 | 99.5953 | 99.5968 |

**Table 1:** NPCR test results.

| Image | Red | Green | Blue |
|---|---|---|---|
| Peppers | 33.4947 | 33.4997 | 33.4933 |
| Lena | 33.4945 | 33.4815 | 33.4178 |
| Sailboat | 33.4420 | 33.4801 | 33.4559 |
| Mandrill | 33.5002 | 33.5688 | 33.5514 |
| Airplane | 33.4530 | 33.4835 | 33.4333 |

**Table 2:** UACI test results.

B. *Histogram Analysis*

An image histogram shows the frequency of the pixel intensity values within the image. Uniform histograms do not provide any useful information, thus to resist statistical attacks, the encrypted image should have uniform histogram as shown in Fig. 4(d) and (h).

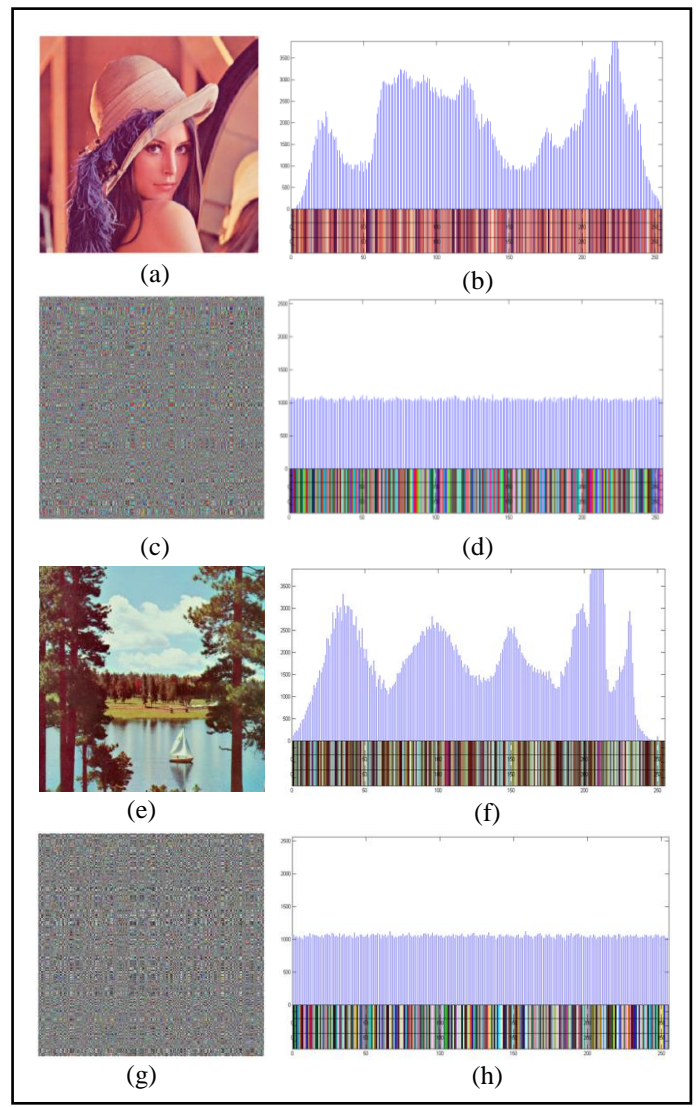

Fig. 4. Histogram analysis: (a) plain text Lena image, (b) histogram of the plain text Lena image, (c) encrypted Lena image, (d) histogram of encrypted Lena image, (e) plain text Sailboat image, (f) histogram of the plain text Sailboat image, (g) encrypted Sailboat image, and (h) histogram of encrypted Sailboat image.

## C. Addjacent Pixel Correlation

The original image pixel correlation is usually very high; a secure image encryption algorithm must produce an encrypted image having low correlation between adjacent pixels to prevent statistical attack.

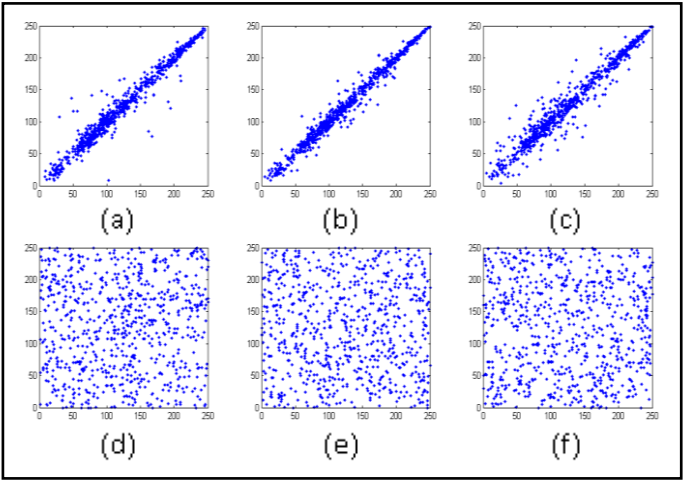

Fig. 5. Adjacent pixel correlation of Lena image: (a) vertical pixel correlation, (b) horizontal pixel correlation, and (c) diagonal pixel correlation of plain text image, (d) vertical pixel correlation, (e) horizontal pixel correlation, and (f) diagonal pixel correlation of encrypted image.

Theresults in Fig. 5 and Table 3 emphasizes that there is no correlationof adjacentpixelsinencryptionimages, indicating that the tested encryption algorithm effectively destroyed the correlation of pixels.

**Table 3:** Correlation coefficients in three directions: horizontal vertical and diagonal.

| Image | | Direction | Red | Green | Blue |
|---|---|---|---|---|---|
| Lena | original | Horizontal | 0.9798 | 0.9691 | 0.9327 |
| | | Vertical | 0.9893 | 0.9825 | 0.9576 |
| | | Diagonal | 0.9697 | 0.9555 | 0.9183 |
| | encrypted | Horizontal | -0.0470 | -0.0377 | -0.0473 |
| | | Vertical | -0.0475 | -0.0501 | -0.0477 |
| | | Diagonal | 0.0015 | 0.0050 | 0.0035 |
| Mandrill | original | Horizontal | 0.9231 | 0.8655 | 0.9073 |
| | | Vertical | 0.8660 | 0.7650 | 0.8809 |
| | | Diagonal | 0.8543 | 0.7348 | 0.8399 |
| | encrypted | Horizontal | -0.0728 | -0.0727 | -0.0691 |
| | | Vertical | 0.0383 | 0.0265 | 0.0359 |
| | | Diagonal | -0.0051 | -0.0017 | -0.0015 |

## D. Informtion Entropy Analysis

The information entropy gives a measure of statistical randomness and unpredictability of information, taken place in encrypted images. The information entropy of a message source $m$ can be described as

$$H(m) = \sum_{i=0}^{2^n-1} p(m_i) \log_2 \left[\frac{1}{p(m_i)}\right] \qquad (6)$$

where $n$ is the number of bits that is required to represent the symbol $m_i$, and $p(m_i)$ is the probability of symbol $m_i$. It is desired from security point of view that encrypted images should have their entropy measure close to the ideal value 8. The entropy measures of the R,B,G components of the original and the encrypted images are shown in Table 4, implies that information leakage in the encrypted images is low and the encrypted image is secure against entropy-base attack.

**Table 4:** Information entropy for plaintext and encrypted images.

| Image | | Red | Green | Blue |
|---|---|---|---|---|
| Peppers | original | 7.3388 | 7.4963 | 7.0583 |
| | encrypted | 7.9992 | 7.9993 | 7.9993 |
| Lena | original | 7.2531 | 7.5940 | 6.9684 |
| | encrypted | 7.9992 | 7.9993 | 7.9993 |
| Sailboat | original | 7.3124 | 7.6429 | 7.2136 |
| | encrypted | 7.9993 | 7.9993 | 7.9993 |
| Mandrill | original | 7.7067 | 7.4744 | 7.7522 |
| | encrypted | 7.9993 | 7.9993 | 7.9993 |
| Airplane | original | 6.7178 | 6.7990 | 6.2138 |
| | encrypted | 7.9993 | 7.9994 | 7.9993 |

## E. Noise and Data Loss Analysis

The strength of the studied encryption algorithm against attacks that may lead to data loss, can be examined by subjecting the encrypted image to noise or pixel cropping attack of some level and then visually observing the quality of the reconstructed images with respect to the original image.

To evaluate the resistance of the encryption algorithm peak signal to noise ratio (PSNR), can be employed.

$$\text{PSNR} = 10 \log_{10}\left(\frac{255^2}{MSE}\right) \qquad (7)$$

$$\text{MSE} = \frac{1}{M \times N} \sum_{i,j} |I_1(i,j) - I_2(i,j)|^2 \qquad (8)$$

where MSE is the mean square error between the recovered image $I_2(i,j)$ and the original image $I_1(i,j)$, and $M$ and $N$ are the rows and columns which represent the width and height of the image.

**Table 5:** PSNR results for speckle noise attack.

| Test | | PSNR | | |
|---|---|---|---|---|
| Image | $\alpha$ | Red | Green | Blue |
| Peppers | 0.05 | 11.6135 | 7.5505 | 11.7968 |
| | 0.1 | 11.2445 | 7.5345 | 11.4232 |
| | 0.3 | 10.4170 | 7.5283 | 10.2242 |
| | 0.5 | 10.0416 | 7.5367 | 9.6083 |
| Lena | 0.05 | 10.8025 | 10.2446 | 13.6931 |
| | 0.1 | 10.4866 | 10.0331 | 13.1787 |
| | 0.3 | 9.5731 | 9.5299 | 11.8703 |
| | 0.5 | 9.1468 | 9.2888 | 11.1701 |
| Mandrill | 0.05 | 10.0970 | 11.2275 | 9.3268 |
| | 0.1 | 9.9166 | 10.9530 | 9.2152 |
| | 0.3 | 9.5018 | 10.2608 | 8.9420 |
| | 0.5 | 9.2917 | 9.9344 | 8.8066 |

Applying speckle noise with densities $\alpha = 0.05, 0.1, 0.3,$ and $0.5$ to the encrypted image, it can be observed in Fig. 6 that the reconstructed image can still be readable with PSNR results listed in Table 5.

Moreover, if the encrypted image is attacked by a data cut of sizes shown in Table 6, then the decryption process is applies to these cropped images. The decrypted images contain most of the visual information as shown in Fig. 7. Thus, the proposed encryption algorithm has excellent performance in noise and data loss attacks.

**Table 6:** PSNR results for data loss attack.

| Test Image | Per. | PSNR | | |
|---|---|---|---|---|
| | | Red | Green | Blue |
| Peppers | 5% | 11.9657 | 7.7292 | 11.8259 |
| | 10% | 11.7077 | 7.7454 | 11.5457 |
| | 20% | 11.5124 | 7.7774 | 10.7218 |
| | 50% | 10.2513 | 7.6863 | 9.3865 |
| Lena | 5% | 11.2526 | 10.3184 | 13.9350 |
| | 10% | 11.0161 | 10.2432 | 13.6034 |
| | 20% | 10.1572 | 10.2377 | 13.0271 |
| | 50% | 9.0974 | 9.5211 | 11.4750 |
| Mandrill | 5% | 10.1306 | 11.3380 | 9.1982 |
| | 10% | 10.0704 | 11.2090 | 9.1950 |
| | 20% | 9.8815 | 10.9865 | 9.1244 |
| | 50% | 9.5040 | 10.2031 | 8.8019 |

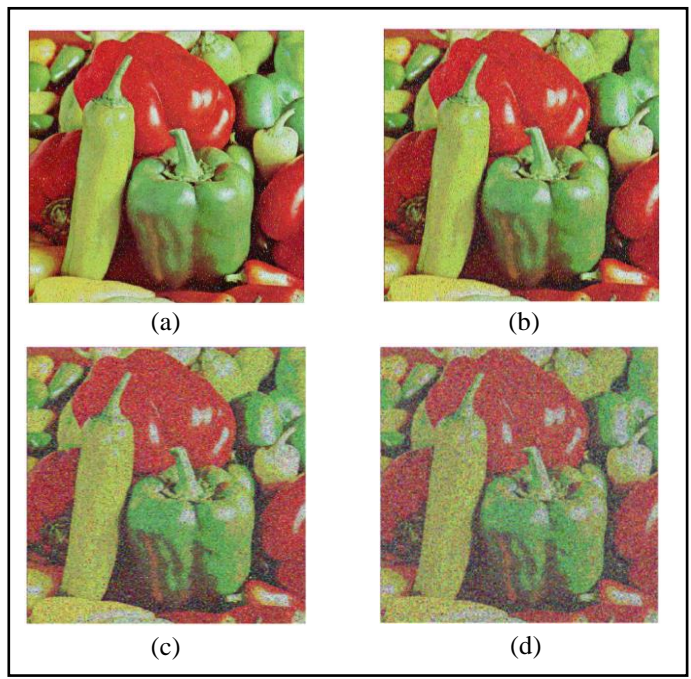

Fig. 6. Speckle noise attack: (a) $\alpha = 0.05$, (b) $\alpha = 0.1$, (c) $\alpha = 0.3$, and (d) $\alpha = 0.5$.

### F. Key Space and Key Sensitivity

The studied image encryption method has two initial parameters (x,y) and iteration N, taking the computational precision to be 10-10, the key space was found to be $(10^{10})^8 = 10^{80}$, which is almost equivalent to $2^{265}$ and that shows that the algorithm is of sufficient key space to resist the exhaustive searching and brute-force attack.

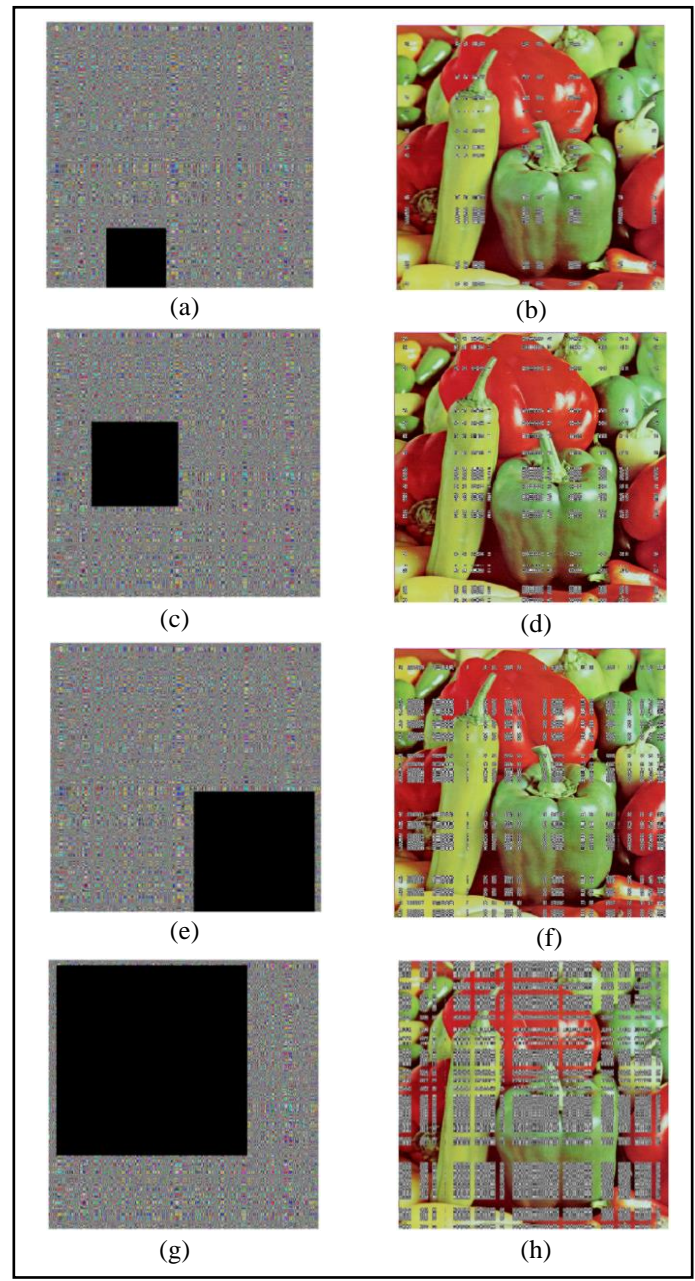

Fig. 7. Data loss attack: (a) encrypted with 5% loss, (b) decrypted with 5% loss, (c) encrypted with 10% loss, (d) decrypted with 10% loss, (e) encrypted with 20% loss, (f) decrypted with 20% loss, (g) encrypted with 50% loss, and (h) decrypted with 50% loss.

### G. Speed Test

For the proposed algorithm time consumption include key stream generation, blocks permutation, and diffusion is measured. The speed analysis test has been carried out on an Intel(R) Core(TM)i7-2670QM 2.2GHz, and 8GB RAM running on Windows 7 and MATLAB R2015a.Thecomputational time for encrypting $512 \times 512 \times 3$ color image with different chaotic maps is 0.4253 second.

## V. COMPARSION WITH OTHER SCHEMES

In this section, we compare our algorithm with other recently proposed algorithms. We see from Table 7 that our scheme is the fastest besides that it exhibits a very large key space, a very good information entropy, and an acceptable UACI and NPCR values as compared with other works.

Table 7: Comparison of someencryption schemes using Lena as test image.

| Metrix | Our | [4] | [5] | [6] | [7] |
|---|---|---|---|---|---|
| Key space | $2^{265}$ | $2^{113}$ | $2^{106}$ | $2^{340}$ | $2^{298}$ |
| Encryption time (s) | 0.425 | 1.853 | 1.82 | 3.76 | 4.9 |
| Entropy | 7.99 | 7.99 | 7.99 | 7.99 | 7.99 |
| NPCR | 99.62 | 99.6 | - | 99.71 | 99.61 |
| UACI | 33.46 | 33.4 | - | 33.45 | 33.46 |

## VI. CONCLUSIONS

In this paper, an image encryption method based on block permutation and XOR operation is introduced. The process involves dividing the image into blocks and then shuffling them. Pixels of the blocks are XORed with the chaotic key stream to get the encrypted image. Simulation results and performance analysis show that the presented algorithm can resist several known attacks effectively, assuring safety performance and secure image encryption. Adjacent pixel correlation of encrypted image is found to be less than that of the original image. Also the results of NPCR and UACI values are very close to the ideal values, which clearly show that the proposed encryption method produces good results.